# Analysis choices can produce apparent rich-club differences: an Alzheimer's disease EEG case study

Kai-Wayne Liu[a,b,†], Ryan Tian En Pang[a,c,†], Pholpat Durongbhan[a,d,e,*]

[†] These authors contributed equally.
[a] Australian Medical Student Research Alliance, Australia
[b] Melbourne Medical School, The University of Melbourne, Parkville, Australia
[c] School of Clinical Sciences at Monash Health, Monash University, Clayton, Australia
[d] Department of Biomedical Engineering, The University of Melbourne, Parkville, Australia
[e] Department of Mechanical Engineering, Graduate School of Engineering, Kyushu University, Fukuoka, Japan

* Corresponding author: Pholpat Durongbhan, p.durongbhan@kyudai.jp, Department of Mechanical Engineering, Graduate School of Engineering, Kyushu University, Fukuoka 819-0395, Japan.

E-mail addresses and ORCID iDs:
kaiwayne.liu@student.unimelb.edu.au, 0000-0002-4496-7921 (K.-W. Liu);
ryanpangte@gmail.com, 0009-0003-6460-1171 (R.T.E. Pang);
p.durongbhan@kyudai.jp, 0000-0002-5164-3482 (P. Durongbhan).

## Abstract

**Background:** The weighted rich-club coefficient tests whether a network's most strongly connected nodes are preferentially interconnected. Its interpretation depends on two often-implicit choices: how the randomised comparison networks are constructed, and whether the hubs are re-identified in each comparison network or held fixed at those of the observed network. When hubs are defined by connection strength, these choices change what the coefficient tests. **New method:** We examined how these choices affect inference using resting-state EEG from 20 people with Alzheimer's disease and 20 controls. Holding the data fixed, we compared four analysis pipelines, crossing two ways of building the comparison networks (global weight reshuffle or strength-conditioned null) with the two hub choices, and ran a calibration study using data constructed to contain no preferential hub coupling beyond nodal strength. **Results:** An apparent eyes-open group difference in rich-club organisation appeared only under the pipeline combining a global weight reshuffle with re-identified hubs. Under the empirical resampling ensemble, that pipeline reported a group difference in 43% of pseudo-studies, always higher in the Alzheimer's disease group, while the other three stayed near the expected 5%. **Comparison with existing methods:** Unlike global weight reshuffling, strength-conditioned nulls approximately preserve nodal strengths; fixing observed hubs also retains the identity of the core being tested. **Conclusions:** The same data can support or dissolve an apparent rich-club difference depending on these choices. No single choice is universally correct, but reporting which were made, and why they fit the question being asked, is essential to interpreting the result.

## 1 Introduction

Alzheimer's disease is increasingly conceptualised as a disorder of large-scale brain network organisation rather than isolated regional degeneration (Delbeuck et al., 2003; Pievani et al., 2011). Distributed cortical communication depends on hub regions: nodes with high connectivity and central topological position (Sporns et al., 2007; van den Heuvel and Sporns, 2011). A hub, in plain terms, is a channel strongly connected to many others; the core, or rich club, is the set of these hubs taken together. When hubs are more strongly interconnected with one another than their individual connectedness alone would predict, a network is said to exhibit rich-club organisation (van den Heuvel and Sporns, 2011). Disruption of this core has been proposed as a candidate mechanism of cognitive decline in Alzheimer's disease (AD) (Daianu et al., 2013).

Studies of rich-club organisation in Alzheimer's disease, however, report conflicting findings. Several structural connectivity studies describe reduced connectivity among hub regions and degraded core efficiency (Daianu et al., 2013; Yan et al., 2018), whereas others report a relatively preserved structural core with disruption concentrated in peripheral regions, and even a higher normalised rich-club coefficient in Alzheimer's disease (Daianu et al., 2015). Across the Alzheimer's disease literature the picture is similarly mixed, with disrupted, preserved, or enhanced core connectivity reported depending on disease stage, modality, and analytical choices (Ma et al., 2022; Yan et al., 2018).

Two explanations are possible: the disease may not produce a consistent change in rich-club organisation, or differences in measurement and analysis may yield different apparent results from similar data. The latter is not hypothetical. In M/EEG functional-network studies, the choice of connectivity metric alone can change whether an Alzheimer's disease effect is detected and in which direction (Schoonhoven et al., 2022), showing that analytical choices can drive the apparent result. This paper concerns that second possibility, and in particular how it applies to the rich-club coefficient, a quantity whose value depends directly on such choices.

The weighted rich-club coefficient is not an absolute quantity: it is defined relative to a set of randomised comparison networks (a null ensemble) representing what the network would look like with no special hub-to-hub coupling (Opsahl et al., 2008; van den Heuvel and Sporns, 2011). What counts as a fair comparison depends on how hubs are defined, by degree (how many connections a node has) or by strength (the total weight of its connections). When hubs are strength-defined, as is common in weighted functional networks, the null must condition on the nodal-strength sequence to separate genuine preferential coupling among hubs from the fact that strong nodes simply carry more weight (Opsahl et al., 2010, 2008); this conditioning is often left implicit.

The issue is distinct from degree-defined structural rich-club analyses, where degree-preserving nulls address a different question. A second choice is equally consequential and less often stated: whether the set of hubs is re-identified inside each randomised network, or held fixed at the hubs of the observed network. These answer different questions. Re-identifying hubs in each surrogate asks whether whichever nodes are strongest tend to interconnect (self-selected rich-clubness); fixing the observed hubs asks whether the nodes strongest in the real data couple preferentially beyond what their strength implies (observed-hub preferential coupling). Neither is incorrect, but they are not interchangeable, and pairing a strength-defined hub set with a null that alters the strength sequence and re-identifies hubs mixes the two.

This work treats these two choices as the object of study. Using a previously described resting-state EEG cohort as a worked example, we hold the data fixed and vary only the analysis, crossing two null models (a global weight reshuffle and an approximately strength-preserving null) with two hub-selection rules (hubs re-identified in each surrogate, or fixed at the observed hubs) to give four specifications. We ask whether an apparent group difference between control and Alzheimer's disease cohorts is stable across these four specifications when they are applied to the same networks, and we then use a calibration study to quantify how often each specification reports a group difference when none is present beyond the nodal-strength structure.

## 2 Materials and methods

### 2.1 Study design and dataset

This study is a secondary analysis of resting-state EEG recordings from 20 healthy control participants (HC group) and 20 participants diagnosed with Alzheimer's disease (AD group). The dataset has been previously described and analysed for pairwise classification features (Durongbhan et al., 2019) and for linear and nonlinear functional connectivity estimation (Zhao et al., 2020).

### 2.2 Participants

Participants were healthy individuals or patients diagnosed with Alzheimer's disease, all of whom had undergone detailed neuropsychological testing and structural and functional brain imaging. All subjects were recruited from Sheffield Teaching Hospitals NHS Trust memory clinic; HC participants were enrolled through opportunity sampling over one year (February 2015 to 2016). Twenty HCs (10 aged <70 years, 11 female; 10 aged ≥70 years; mean age 67 ± 12 years) and 20 ADs (10 female; 16 aged <70 years; 4 aged ≥70 years; mean age 64 ± 8 years) were included. Alzheimer's disease diagnosis followed NINCDS-ADRDA criteria (Dubois et al., 2007), based on consensus incorporating clinical history, neurological examination, neuropsychological scores, and neuroradiological findings. Demographic, neuropsychological, and structural MRI details are reported in prior work (Blackburn et al., 2018; Durongbhan et al., 2019). The study was approved by the Yorkshire and the Humber (Leeds West) Research Ethics Committee (reference 14/YH/1070), and all participants provided informed consent.

### 2.3 EEG recordings and epoch-wise network construction

#### 2.3.1 EEG acquisition and preprocessing

Recordings were acquired with an XLTEK 128-channel headbox at 2 kHz using an earlobe reference, over 30-minute resting-state sessions with distinct eyes-open and eyes-closed segments. For each participant and state, the first artefact-free 12-second epoch was selected to avoid selection bias; a 2 Hz high-pass and a notch filter were applied, and the middle 4-second segment was used for connectivity analysis to avoid boundary effects (Zhao et al., 2020). Twenty-three bipolar channels were analysed (Supplementary Table S1).

#### 2.3.2 Functional connectivity estimation

Pairwise connectivity was estimated using the error reduction ratio model (Zhao et al., 2020), whose linear and nonlinear components are symmetric, bidirectional measures usable directly as undirected edge weights. The present analysis uses the linear component, with eyes-open segments as primary and eyes-closed reported alongside; the linear component was chosen as a working demonstration because it is larger in magnitude than its nonlinear counterpart. Channel pairs sharing an electrode were excluded, yielding 207 permissible edges. This permissible-edge set is uneven across nodes, since bipolar derivations share electrodes with differing numbers of other derivations; the resulting fixed set of edges, referred to here as the montage mask, is preserved throughout the analysis.

2.3.3 Epoch-wise network construction

Each 4-second segment was divided into five consecutive non-overlapping mini-epochs of 0.8 seconds (Zhao et al., 2020), and connectivity was estimated within each. For every channel pair, the edge weight is the linear connectivity value averaged across these five mini-epochs, giving one 23 by 23 weighted, undirected network per participant and state.

**2.4 Weighted rich-club coefficient and hub definition**

Node strength was the sum of incident edge weights (Rubinov and Sporns, 2010). Hubs were the top-$k$ nodes by strength (primary $k$ = 5, the top ~20% of 23; robustness across $k$ = 4 to 9; Supplementary Table S2). The weighted rich-club coefficient (Opsahl et al., 2008) is the sum of edge weights among the hub set divided by the sum of the equally many strongest permissible edge weights in the network.

Because the montage mask is uneven, the number of permissible edges among a hub set depends on which nodes are hubs. To keep the denominator constant across observed and null networks, the fixed-hub analysis proceeds as follows. For each observed network: (i) identify the top-five hubs by observed nodal strength; (ii) count the permissible edges among that hub set, $E_{obs}$; (iii) retain both the observed hub set and $E_{obs}$ across every null network generated for that observed network; (iv) compute the denominator as the sum of the $E_{obs}$ strongest permissible edge weights in each network. The hub set and edge count are thus held fixed, so observed and null coefficients are directly comparable. In the re-selected-hub procedure, the top-five hubs and edge count are instead recomputed within each null network, defining the different estimand described in Section 1.

**2.5 Null models and analysis specifications**

Two null models were used. The first is a mask-constrained global weight reshuffle: observed edge weights are permuted among the permissible edges, preserving the weight distribution and the montage mask but not the nodal-strength sequence. The second is an approximately strength-preserving, mask-constrained null: edge weights are redistributed while holding each node's strength close to its observed value (mean per-node strength error 1.9%, and not different between groups; Supplementary Table S3; generation procedure listed in Supplementary Section S4). Each null was crossed with each hub-selection rule (fixed or re-selected), giving four specifications. For every empirical coefficient, 1,000 null networks were generated, enough to give a stable normalising mean while remaining tractable across the full analysis and calibration; the coefficient was normalised by the mean of the null distribution.

**2.6 Calibration under a no-extra-core ground truth**

To quantify how often each specification reports a group difference when none is present beyond nodal strength, we generated pseudo-studies under a ground truth containing no preferential hub coupling. For each pseudo-study, a pseudo-network was generated for every participant and state by resampling that participant's observed network under the strength-preserving construction, which retains the participant's channel-indexed strength sequence, including any between-group strength difference, but introduces no hub-to-hub coupling beyond what that strength structure implies. Pseudo-networks were generated per mini-epoch and averaged across the five mini-epochs, as in the primary analysis, and the original group labels were retained. All four specifications were then applied. Across 2,000 valid pseudo-studies (no exclusions), we report, for each specification, the proportion returning a group difference at $\alpha = 0.05$ (the conditional overcall rate), Wilson binomial 95% confidence intervals, and the proportion of those rejections that were higher in the AD group. To test whether these rates depend on how the pseudo-networks are generated, the analysis was repeated with a different strength-preserving generator that redraws edge weights before matching each node's target strength (randomised Sinkhorn; Supplementary Table S5; generation procedure listed in Supplementary Section S6).

## 2.7 Statistical analysis

Between-group comparisons used the Mann-Whitney U test, with the rank-biserial correlation r as the effect size. A negative r indicates higher values in the AD group; for readability we describe such differences as higher in the AD group rather than by the sign of r. Tests are exploratory and reported without correction; the focus is the pattern across specifications rather than any single test. Rank-biserial effect sizes are reported with bootstrap 95% confidence intervals (10,000 resamples within group). Analyses were implemented in Python 3. To ensure reproducibility, every randomised step uses a fixed seed, and the main analyses derive theirs from a single base seed and the identity of the analysis cell. The analysis code was developed with the assistance of a generative AI tool (Claude, Anthropic, USA), reviewed by the authors, and publicly released in full.

# 3 Results

## 3.1 The apparent group difference is specification-dependent

Under the four analysis specifications, the eyes-open group comparison at $k = 5$ diverged sharply (Table 1). A group difference in the normalised rich-club coefficient was detected only under the global weight-reshuffle null combined with null-wise hub re-selection (AD 1.09, HC 0.93, $p = 0.034$, $r = -0.40$). No group difference was detected under the three alternative specifications: global reshuffle with fixed observed hubs ($p = 0.78$, $r = -0.06$), strength-conditioned null with re-selected hubs ($p = 0.54$, $r = -0.12$), and strength-conditioned null with fixed observed hubs ($p = 0.47$, $r = -0.14$). The same pattern held across club sizes: the eyes-open difference was present only in the reshuffle-plus-re-selection specification, at $k$ = 5, 7, 8 and 9 ($k = 6$, $p = 0.053$), and was absent in every eyes-closed comparison and under every correctly matched specification (Supplementary Table S2). A significant result in eyes-open recordings and a non-significant result in eyes-closed recordings do not, by themselves, establish a state-dependent group effect. The calibration shows that the eyes-open contrast can arise under the no-extra-core reference ensemble. No AD-HC difference in normalised observed-hub coupling was detected under either fixed-hub analysis.

| **Null model** | **Hub selection** | **AD ($\varphi_{norm}$)** | **HC ($\varphi_{norm}$)** | $p$ | $r$ **[95% CI]** |
|---|---|---|---|---|---|
| Global weight reshuffle | Re-selected in surrogates | 1.09 | 0.93 | 0.034 | −0.40 [−0.72, −0.03] |
| Global weight reshuffle | Fixed (observed) | 1.96 | 1.91 | 0.776 | −0.06 [−0.42, 0.31] |
| Strength-conditioned | Re-selected in surrogates | 1.13 | 1.09 | 0.543 | −0.12 [−0.50, 0.26] |
| Strength-conditioned | Fixed (observed) | 1.13 | 1.10 | 0.473 | −0.14 [−0.51, 0.23] |

*Table 1: Normalised weighted rich-club coefficient by group under four analysis specifications (eyes-open linear connectivity, $k$ = 5). A group difference was detected only when a global weight-reshuffle null was combined with null-wise hub re-selection. $r$ is the rank-biserial effect size, with its bootstrap 95% confidence interval in brackets; a negative $r$ indicates higher values in the AD group. Coefficients are normalised to specification-specific null ensembles and are not comparable across rows; the relevant comparison is between groups within each row.*

### 3.2 Conditional calibration of the four specifications

Under the no-extra-core ensemble (Section 2.6), the four specifications behaved very differently (Figure 1). The global-reshuffle-plus-re-selection specification returned a group difference in 43.2% of the 2,000 pseudo-studies (95% CI 41.0 to 45.4), every one of them higher in the AD group. The three alternative specifications were approximately calibrated at the nominal level within that empirical resampling ensemble, with conditional overcall rates of 2.4% (reshuffle, fixed hubs), 4.5% (strength-conditioned, re-selected hubs), and 4.3% (strength-conditioned, fixed hubs), and no directional bias (53 to 55% in the AD direction). The observed eyes-open effect ($r$ = −0.40) fell at the 36th percentile of the distribution of pseudo-study effects produced by the reshuffle-plus-re-selection specification, and was therefore not unusual under the no-extra-core ensemble. The high conditional overcall rate was not reproduced by a stochastic generator that redrew edge weights before matching nodal strengths (Supplementary Table S5); this dependence on the generating model is considered in the Discussion.

Surrogate-level re-selection also interacts with the uneven montage. Under the global reshuffle a node's expected surrogate strength is proportional to its number of permissible edges, and the re-selected hub sets followed that count (Spearman $\rho$ = 0.96), recovering on average 1.2 of the 5 observed hubs, similarly in both groups; under the strength-conditioned null they recovered 4.8 of 5 (Supplementary Table S7). The comparison coefficient of the mismatched specification is therefore computed for a mask-determined node set rather than the observed one.

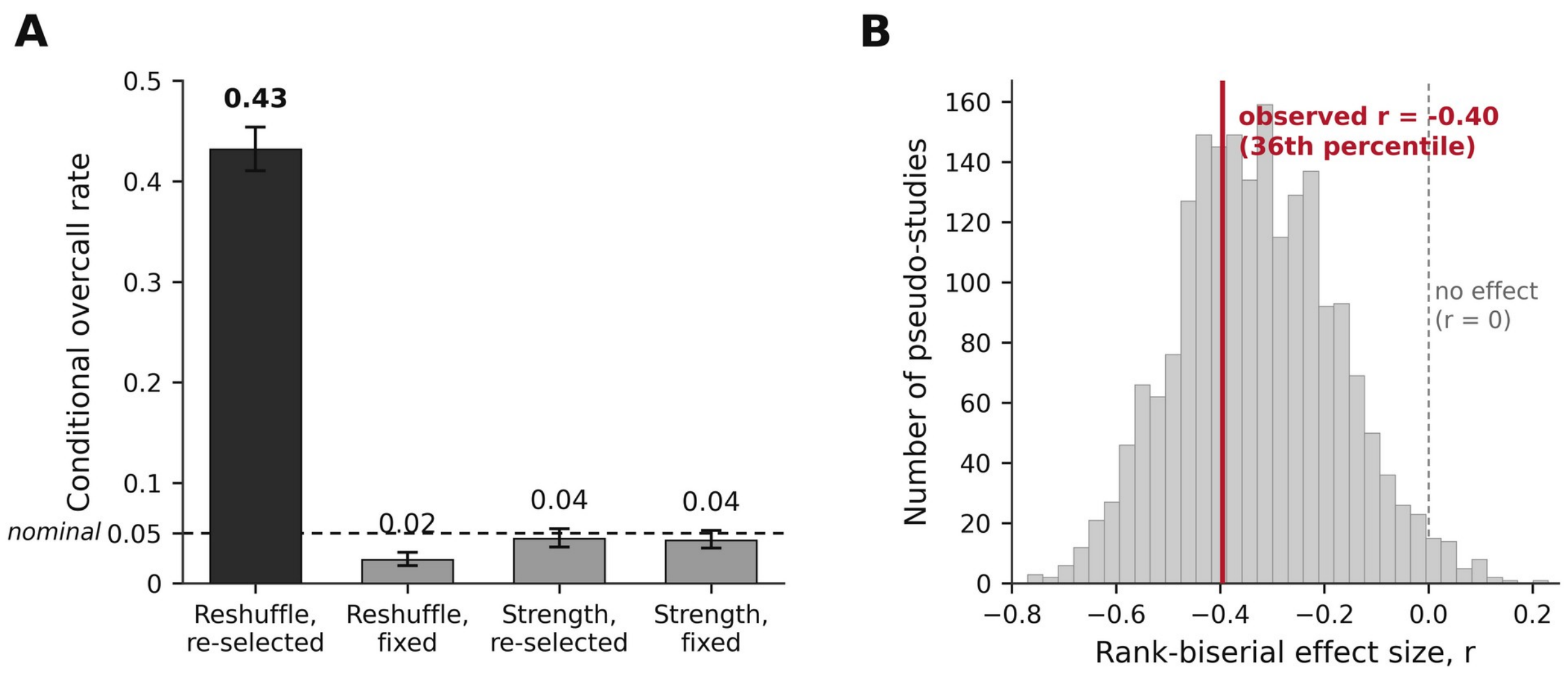

*Figure 1: Calibration of the four analysis specifications under a no-extra-core ground truth. Pseudo-networks were drawn from an approximately strength-preserving resampling of each participant's observed network, preserving the channel-indexed nodal-strength structure (including the between-group strength difference) but containing no hub-to-hub coupling beyond that strength structure; 2,000 pseudo-studies were generated. The four specifications cross two null models (global weight reshuffle, or strength-conditioned) with two hub-selection rules (hubs re-selected in each surrogate, or fixed at the observed hubs); panel labels abbreviate these as Reshuffle/Strength and re-selected/fixed. (A) Conditional overcall rate (proportion of pseudo-studies returning a group difference at $\alpha = 0.05$) for each specification, with Wilson binomial 95% confidence intervals; the dashed line marks the nominal 0.05 level. The reshuffle, re-selected specification returned a group difference in 43% of studies; the other three were approximately calibrated. (B) Distribution of pseudo-study rank-biserial effects ($r$) for the reshuffle, re-selected specification; negative $r$ indicates higher values in the AD group. The observed eyes-open effect ($r = -0.40$, red line) falls at the 36th percentile, and all of the significant pseudo-study effects under this specification were higher in the AD group.*

### 3.3 Supporting analyses do not indicate preferential observed-hub coupling

Several further analyses were consistent with the fixed-hub result. The internal hub-strength fraction, the proportion of the connectivity carried by the five strongest nodes that is directed to other strong nodes, did not differ between groups (AD 0.27, HC 0.26), although the total connectivity carried by those nodes was directionally higher in the AD group, consistent with the raw connectivity pattern rather than with preferential core coupling (Supplementary Table S8). A test of relative core enrichment, the difference between core-core and other edge classes after normalising each participant's weights by their network mean, showed no group difference (Supplementary Table S9). The fixed-hub result was stable across alternative hub definitions, including hubs ranked by mean permissible-edge weight and by strength residualised on each node's number of permissible edges; the re-selection-based contrast attenuated when hub ranking was adjusted for permissible-edge opportunity (Supplementary Table S10). Mean eyes-open linear connectivity was higher in the AD group, but the group-by-state comparison of the eyes-closed-to-eyes-open change was null, so this difference is not interpreted as state-dependent (Supplementary Table S11). Taken together, these analyses point to elevated overall connectivity in the AD group rather than preferential coupling within the hub core.

## 4 Discussion

This study shows that an apparent rich-club difference between groups can be a property of the analysis rather than of the data. In the same Alzheimer's disease EEG networks, an eyes-open difference in weighted rich-club organisation appeared only when a global weight-reshuffle null was combined with null-wise hub re-selection, and disappeared once coupling among the observed strength-defined hubs was assessed against a strength-conditioned null. The supporting analyses found no greater preferential allocation of connectivity to the hub core in the AD group, so the difference seen under the first specification cannot be read as evidence of a core-specific effect. The supporting decomposition points to a strength-driven rather than a preferential-coupling interpretation: a strength-conditioned null accounts for the greater total connectivity the strength-defined hubs carry, whereas a weight-reshuffle null does not (Opsahl et al., 2008). The practical consequence is direct: unless the null model and the hub-selection rule are chosen to match the question, a weighted rich-club analysis can report elevated hub strength as though it were preferential coupling within the core.

The four specifications disagree because they measure different things. The normalised coefficient is a ratio (Opsahl et al., 2008). The numerator comes from the observed network and the denominator comes from the surrogates, so the null model and the hub-selection rule decide what the ratio means (Rubinov and Sporns, 2010; van den Heuvel and Sporns, 2011). With a strength-conditioned null and the observed hubs held fixed, the ratio tests whether the observed hubs are more strongly connected to each other than their strengths and the montage would predict. Re-selecting hubs under the same null gives almost the same test, because the surrogates recover 4.8 of the five observed hubs on average (Supplementary Table S7); this is why those two rows of Table 1 agree. The global reshuffle removes the strength sequence. When hubs are also re-selected, it removes the hub identity as well: a node's expected surrogate strength is proportional to its number of permissible edges, so the surrogate hubs are set by the montage rather than by the data, and they share on average 1.2 of 5 nodes with the observed hubs (Supplementary Table S7). The denominator of the mismatched specification is therefore the coupling of a mask-determined node set in networks that no longer carry the participant's strength profile. A uniform difference in overall connectivity cannot by itself change the ratio, since scaling every weight scales numerator and denominator alike; what the mismatched specification is exposed to is the strength and edge-weight structure that its surrogates discard. The calibration reproduces the overcall from a ground truth that keeps each participant's strength sequence and weight distribution but adds no hub-to-hub coupling, and the randomised Sinkhorn generator, which keeps the strength sequence alone, does not (Supplementary Table S5); the present analyses do not identify which retained feature drives it.

The mismatched specification was markedly anti-conservative: under a ground truth with no core structure beyond nodal strength, it reported a core-specific group difference in 43% of cases, always in the AD direction, while the three matched specifications stayed near the nominal rate. Within the empirical resampling ensemble, pairing an unconditioned null with hub re-selection was prone to reporting strength-associated differences as core-specific, whereas conditioning the null on strength and fixing the observed hubs was not. The hub-selection rule matters as much as the null, and the two must be chosen together (Rubinov and Sporns, 2010).

The mechanism does not depend on the connectivity metric. It needs three things: hubs defined by strength, a null that does not preserve the strength sequence, and hub re-selection inside the surrogates. Any weighted connectivity measure with unequal nodal strengths supplies the first; the weighted phase lag index (wPLI), coherence and correlation-based matrices all do. The other two are choices made by the analyst. The size of the effect does not transfer. The conditional overcall rate depends on the weight structure of the data, and the two ground truths used here show how much (43% under one, 2% under the other). Dense matrices with no montage mask lose the mask-driven part of re-selection but keep the strength-driven part. Binarising a matrix makes hubs degree-defined, and a degree-preserving null then returns the same hubs in every surrogate, so the mismatch cannot arise; thresholding without binarising keeps the exposure if hubs remain strength-defined. The same reasoning covers other modalities. The canonical structural rich club defines hubs by degree and normalises against degree-preserving surrogates (Daianu et al., 2015, 2013; van den Heuvel and Sporns, 2011), so its hub set is the same in every surrogate and the mismatch does not occur there. Weighted structural or fMRI networks with strength-defined hubs are exposed to it. The calibration transfers to any non-negative weighted network on a fixed node set, with the montage mask replaced by the network's own edge set; signed matrices, such as correlation matrices with negative entries, need their positive and negative weights treated separately.

Several limitations apply. The cohort is a single previously described sample of modest size (Blackburn et al., 2018; Durongbhan et al., 2019), used here as a worked example rather than for clinical inference. The size of the 43% overcall rate is specific to the ground truth used: it was obtained under a resampling that preserves each participant's observed strength sequence and weight structure, and was not reproduced by a generator that redraws edge weights before matching strengths, so the exact figure depends on the generating model and we do not identify which retained property is responsible. The strength-preserving null is itself approximate rather than exact. Electrode placement varies subtly between participants. All four specifications were applied to the same measured networks, so the comparison between them is conditional on those networks; placement variability, and any interaction between it and the specification choices, was not quantified separately, and it adds noise that may dilute a real effect in a cohort of this size. Finally, the demonstration covers one connectivity measure and one family of strength-based hub definitions, and the interaction between these specification choices and the choice of connectivity metric (Schoonhoven et al., 2022) remains to be measured rather than inferred.

The specification effect demonstrated here is a property of weighted rich-club inference rather than of this dataset. A natural extension of this work is to characterise which features of the empirical weight structure drive the overcall, and to establish whether the effect persists for nonlinear connectivity and for other hub definitions. More broadly, the calibration approach used here, applying candidate pipelines to data built to contain no effect, is a general way to check whether a network statistic is reporting structure or a product of the analysis. This mirrors a broader concern in quantitative imaging, where the processing definition applied to identical data can change the direction of an apparent group difference (Durongbhan et al., 2026).

## 5 Conclusion

Whether an apparent rich-club difference reflects biology or analysis depends on two choices that are easy to leave implicit: how the comparison networks are built, and whether hubs are re-identified or held fixed. In this Alzheimer's disease EEG cohort, an apparent group difference was present under one combination of these choices and absent under the others, and a calibration analysis showed, within that empirical resampling ensemble, that the combination producing it reports a difference roughly 43% of the time when none is present. No single choice is universally correct, but which choices were made, and why they fit the question, should be stated explicitly. Reported this way, a rich-club analysis is far less likely to have a difference driven by hub strength mistaken for preferential coupling within the core.

**Acknowledgement**
We thank Tony Bohan Huang for facilitating the collaboration that initiated this project.

**Author contribution**

**Kai-Wayne Liu:** Conceptualisation, Formal analysis, Investigation, Methodology, Software, Writing – original draft, Writing – review & editing.

**Ryan Tian En Pang:** Conceptualisation, Methodology, Writing – original draft, Writing – review & editing.

**Pholpat Durongbhan:** Conceptualisation, Data curation, Formal analysis, Investigation, Methodology, Project administration, Resources, Supervision, Validation, Visualisation, Writing – original draft, Writing – review & editing.

**Role of the funding source**
This research did not receive any specific grant from funding agencies in the public, commercial, or not-for-profit sectors.

**Declaration of competing interests**

All authors declare that they have no competing interests.

**Declaration of generative AI use during manuscript preparation**

ChatGPT (5.5, 5.6 Sol, 5.6 Astra; OpenAI, USA) and Claude (Opus 4.8, 5.0, 5.5 and Fable 5.1; Anthropic, USA) were used to improve grammatical clarity and flow, and Claude (Opus 5.5 and Fable 5.1) was used to support code development, debugging and refactoring. All generative AI tools were accessed in May to July and September 2026. After using these tools, the authors reviewed and edited the content as necessary and take full responsibility for the content of the published article.

**Data availability**

The data that support the findings of this study are publicly available within the accompanying software repository.

**Code availability**

The analysis code, the derived connectivity matrices, a one-command reproduction script and the output tables are available at https://github.com/PDurongbhan/rich-club-specification-eeg and archived at Zenodo (https://doi.org/10.5281/zenodo.22983739).

**Supplementary materials for 'Analysis choices can produce apparent rich-club differences: an Alzheimer's disease EEG case study'**

## S1 Montage and permissible-edge layout

Connectivity was computed between 23 bipolar EEG derivations formed from the 10-20 electrode montage. Because two bipolar derivations that share a common electrode are not independent, channel pairs sharing an electrode were excluded, leaving 207 permissible edges. The number of permissible edges is not the same for every derivation, since each shares electrodes with a different number of others; this is the montage mask referred to in the main text. Table S1 lists the 23 derivations and the number of permissible edges incident on each.

| Bipolar derivation | Permissible edges | Bipolar derivation | Permissible edges |
|---|---|---|---|
| **F8-F4** | 20 | **C4-P4** | 17 |
| **F7-F3** | 20 | **C3-P3** | 17 |
| **F4-C4** | 17 | **T4-T6** | 20 |
| **F3-C3** | 17 | **T3-T5** | 20 |
| **F4-FZ** | 18 | **P4-PZ** | 18 |
| **F3-FZ** | 18 | **P3-PZ** | 18 |
| **FZ-CZ** | 17 | **T6-O2** | 19 |
| **T4-C4** | 18 | **T5-O1** | 19 |
| **T3-C3** | 18 | **P4-O2** | 18 |
| **C4-CZ** | 16 | **P3-O1** | 18 |
| **C3-CZ** | 16 | **O1-O2** | 18 |
| **CZ-PZ** | 17 | | |

**Table S1:** The 23 bipolar EEG derivations (10-20 montage) used as network nodes, with the number of permissible edges incident on each. Permissible edges exclude pairs of derivations that share a common electrode, giving 207 permissible edges in total. Derivations are listed in anterior-to-posterior order, following Durongbhan et al. (2019) and Zhao et al. (2020). The per-node count ranges from 16 to 20, confirming that the permissible-edge set (the montage mask) is uneven across nodes. This unevenness is why the fixed-hub analysis holds both the observed hub set and the observed permissible-edge count constant when forming the null distribution (main text, Section 2.4): without fixing the edge count, the rich-club denominator would vary with which nodes happen to be selected as hubs.

### S2 Club-size robustness: full specification sweep across $k$ and state

The main text reports the four analysis specifications at the primary club size ($k$ = 5) in the eyes-open state. Table S2 gives the complete sweep: all four specifications, club sizes $k$ = 4 to 9, in both eyes-open and eyes-closed states. This shows the full extent of the specification-dependent effect and confirms it is confined to a single specification in a single state rather than being specific to the primary $k$.

| Specification | $k$ | AD | HC | $p$ | $r$ |
|---|---|---|---|---|---|
| ***Eyes-open*** | | | | | |
| **Reshuffle, re-selected** | 4 | 1.09 | 0.93 | 0.133 | -0.28 |
| **Reshuffle, re-selected** | **5** | **1.09** | **0.93** | **0.034** | **-0.40** |
| **Reshuffle, re-selected** | **6** | **1.09** | **0.97** | **0.053** | **-0.36** |
| **Reshuffle, re-selected** | **7** | **1.10** | **0.98** | **0.014** | **-0.46** |
| **Reshuffle, re-selected** | **8** | **1.09** | **1.01** | **0.036** | **-0.39** |
| **Reshuffle, re-selected** | **9** | **1.09** | **1.00** | **0.007** | **-0.50** |
| **Reshuffle, fixed** | 4 | 2.07 | 2.05 | 0.989 | +0.01 |
| **Reshuffle, fixed** | 5 | 1.96 | 1.91 | 0.776 | -0.06 |
| **Reshuffle, fixed** | 6 | 1.86 | 1.87 | 0.756 | -0.06 |
| **Reshuffle, fixed** | 7 | 1.81 | 1.77 | 0.636 | -0.09 |
| **Reshuffle, fixed** | 8 | 1.73 | 1.74 | 0.989 | +0.01 |
| **Reshuffle, fixed** | 9 | 1.68 | 1.64 | 0.695 | -0.07 |
| **Strength, re-selected** | 4 | 1.14 | 1.14 | 0.882 | +0.03 |
| **Strength, re-selected** | 5 | 1.13 | 1.09 | 0.543 | -0.12 |
| **Strength, re-selected** | 6 | 1.10 | 1.10 | 0.715 | -0.07 |
| **Strength, re-selected** | 7 | 1.10 | 1.09 | 0.543 | -0.12 |
| **Strength, re-selected** | 8 | 1.09 | 1.10 | 0.882 | -0.03 |
| **Strength, re-selected** | 9 | 1.08 | 1.07 | 0.457 | -0.14 |
| **Strength, fixed** | 4 | 1.14 | 1.14 | 0.818 | +0.04 |
| **Strength, fixed** | 5 | 1.13 | 1.10 | 0.473 | -0.14 |
| **Strength, fixed** | 6 | 1.11 | 1.11 | 0.695 | -0.07 |
| **Strength, fixed** | 7 | 1.10 | 1.09 | 0.473 | -0.14 |
| **Strength, fixed** | 8 | 1.09 | 1.10 | 0.968 | -0.01 |
| **Strength, fixed** | 9 | 1.09 | 1.07 | 0.425 | -0.15 |
| ***Eyes-closed*** | | | | | |
| **Reshuffle, re-selected** | 4 | 1.16 | 1.13 | 0.882 | -0.03 |
| **Reshuffle, re-selected** | 5 | 1.16 | 1.15 | 0.925 | -0.02 |
| **Reshuffle, re-selected** | 6 | 1.18 | 1.11 | 0.394 | -0.16 |
| **Reshuffle, re-selected** | 7 | 1.18 | 1.13 | 0.239 | -0.22 |
| **Reshuffle, re-selected** | 8 | 1.16 | 1.12 | 0.379 | -0.17 |
| **Reshuffle, re-selected** | 9 | 1.13 | 1.11 | 0.636 | -0.09 |
| **Reshuffle, fixed** | 4 | 2.01 | 2.14 | 0.273 | +0.20 |
| **Reshuffle, fixed** | 5 | 1.93 | 2.06 | 0.324 | +0.18 |
| **Reshuffle, fixed** | 6 | 1.89 | 1.89 | 0.797 | +0.05 |
| **Reshuffle, fixed** | 7 | 1.83 | 1.86 | 0.756 | +0.06 |
| **Reshuffle, fixed** | 8 | 1.74 | 1.77 | 0.525 | +0.12 |
| **Reshuffle, fixed** | 9 | 1.64 | 1.69 | 0.598 | +0.10 |
| **Strength, re-selected** | 4 | 1.17 | 1.20 | 0.457 | +0.14 |
| **Strength, re-selected** | 5 | 1.15 | 1.19 | 0.441 | +0.14 |
| **Strength, re-selected** | 6 | 1.15 | 1.12 | 0.925 | +0.02 |
| **Strength, re-selected** | 7 | 1.14 | 1.12 | 0.903 | -0.03 |
| **Strength, re-selected** | 8 | 1.11 | 1.10 | 0.925 | +0.02 |
| **Strength, re-selected** | 9 | 1.07 | 1.08 | 0.797 | +0.05 |
| **Strength, fixed** | 4 | 1.17 | 1.22 | 0.394 | +0.16 |
| **Strength, fixed** | 5 | 1.17 | 1.20 | 0.561 | +0.11 |
| **Strength, fixed** | 6 | 1.16 | 1.12 | 0.882 | +0.03 |
| **Strength, fixed** | 7 | 1.14 | 1.13 | 0.946 | -0.01 |
| **Strength, fixed** | 8 | 1.11 | 1.10 | 0.989 | +0.01 |
| **Strength, fixed** | 9 | 1.07 | 1.08 | 0.797 | +0.05 |

**Table S2:** Group comparison of the normalised weighted rich-club coefficient under all four analysis specifications, across club sizes $k$ = 4 to 9, in both states. AD and HC, group mean normalised coefficient; $p$, Mann-Whitney U test $p$-value; $r$, rank-biserial effect size (a negative $r$ indicates higher values in the AD group). Bold rows are significant at $p < 0.05$. A group difference is present only in the eyes-open state under the reshuffle, re-selected specification, where it holds at $k$ = 5, 7, 8 and 9 ($k$ = 6, $p$ = 0.053); it is absent at every $k$ under the other three eyes-open specifications and absent throughout the eyes-closed state. That the difference appears in exactly one of the four specifications, is not reproduced by the two correctly strength-conditioned specifications, and does not appear in the eyes-closed data together indicate that it is a property of the analysis specification rather than a consistent group effect. The primary analysis ($k$ = 5) is therefore representative of the broader pattern rather than a favourable choice of club size.

## S3 Strength-preserving null: validation and group-neutrality

The strength-preserving null redistributes edge weights while holding each node's strength close to its observed value. Two properties matter for the analysis. First, the null should preserve strength closely, so that it genuinely tests preferential coupling rather than a difference in total nodal strength. Second, it should preserve strength equally well in both groups: if the null matched observed strength more accurately in one group than the other, the between-group comparison of normalised coefficients would be confounded by differential null quality rather than reflecting the networks themselves. Table S3 reports the strength-matching error, overall and by group, for the eyes-open networks.

| Group | n | Mean per-node strength error | Group comparison |
|---|---|---|---|
| **All participants** | 40 | 1.89% | - |
| **Alzheimer's disease** | 20 | 1.80% | $p$ = 0.298 |
| **Control** | 20 | 1.97% | |

**Table S3:** Strength-matching error of the approximately strength-preserving null, computed on the eyes-open networks. For each participant, the null was generated 200 times, and the mean absolute difference between null and observed nodal strength was expressed as a percentage of mean observed strength (mean per-node strength error), then averaged across nulls. The overall mean error was 1.89% (individual participants ranged from 1.04% to 3.82%), confirming that the null preserves nodal strength closely. The error did not differ between groups (Alzheimer's disease 1.80%, control 1.97%; Mann-Whitney U test, $p$ = 0.298), so the null matches observed strength equally well in both groups. The between-group comparison of normalised rich-club coefficients is therefore not confounded by differential null quality: any group difference in a normalised coefficient reflects the networks rather than an asymmetry in how well the null reproduces each group's nodal strength.

**S4 Null generation procedure**

Inputs: the participant network $W$ (23 × 23, symmetric), the montage mask of permissible edges (207 in this study), the number of candidate swaps evaluated per iteration $K$ = 64, and the number of passes $P$ = 6. Output: a null network with the same mask and the same multiset of edge weights as $W$, and a nodal-strength sequence close to that of $W$.

1. Let $E$ be the list of permissible edges $(i, j)$ with $i < j$, and $w$ the vector of their observed weights.
2. Assign a random permutation of $w$ to $E$. This is the global weight reshuffle: it preserves the mask and the weight multiset but not the strength sequence.
3. For every node $n$, compute the strength deviation $d_n = s_n(null) - s_n(W)$.
4. Repeat $\lfloor P \cdot \frac{|E|}{K} \rfloor$ times:
    a. Draw $K$ random pairs of edge positions $(a, b)$, with replacement.
    b. For each pair, compute the change in $\sum_n d_n^2$ that swapping the weights of $a$ and $b$ would produce. Only the four end-nodes of the two edges are affected: each gains or loses the weight difference $w_b - w_a$.
    c. Apply, one at a time, every swap whose change is negative and whose two weights still equal the values used to evaluate it (an earlier swap in the same batch may have moved them; such stale proposals are skipped), updating $d$ for the four affected nodes.
5. Return the symmetric matrix formed from $E$ and the final weights.

The procedure is an approximate greedy descent on the squared strength deviation: within a batch, candidate swaps are evaluated against the deviations at the start of the batch, so a swap applied later in the same batch that shares a node with an earlier one is not guaranteed to reduce the objective. The final strength-matching error (Table S3) is the check on the result. It neither adds nor removes edges and never changes the value of a weight, only its position, so the mask and the weight multiset are exactly those of the observed network. In this study it leaves a mean per-node relative strength error of 1.89% (Table S3). This algorithm is implemented in strength_null_batch in fast_null2.py of the released code.

## S5 Alternative generator for the calibration ground truth

## (randomised Sinkhorn)

The calibration in the main text draws pseudo-networks by resampling each participant's observed network, which preserves both the nodal-strength sequence and the observed multiset of edge weights. To test whether the resulting overcall rate depends on that particular construction, the calibration was repeated with a different generator that preserves the nodal-strength sequence but does not retain the observed weight structure. This generator starts each pseudo-network from a random weight matrix (gamma-distributed on the permissible edges) and then rescales it iteratively so that every node matches its target strength. It therefore imposes the same strength sequence, including the between-group strength difference, while drawing the underlying weights afresh each time.

Two checks confirm the generator behaves as intended. Independent draws for the same participant are only weakly correlated (mean pairwise edge-weight correlation 0.25), so it produces genuinely variable pseudo-networks rather than near-copies of a single template. And the strength-matching error is 0.00%, so target strengths (and the group difference in strength) are reproduced essentially exactly. Table S5 reports the resulting conditional overcall rates.

| Specification | Overcall rate | 95% CI | Higher in AD |
|---|---|---|---|
| **Reshuffle, re-selected** | 2.1% | 1.5 to 2.8 | 53% |
| **Reshuffle, fixed** | 1.2% | 0.8 to 1.8 | 57% |
| **Strength, re-selected** | 5.1% | 4.2 to 6.2 | 52% |
| **Strength, fixed** | 4.5% | 3.7 to 5.5 | 54% |

**Table S5:** Conditional overcall rate for each specification under the randomised Sinkhorn generator, which preserves the nodal-strength sequence but redraws the edge weights (2,000 pseudo-studies). Overcall rate, the proportion of pseudo-studies returning a group difference at $p < 0.05$; 95% CI, Wilson binomial interval; Higher in AD, the proportion of significant results in which the coefficient was higher in the Alzheimer's disease group. Under this generator the reshuffle, re-selected specification overcalls in only 2.1% of pseudo-studies, compared with 43% under the resampling generator used in the main text, and it shows no directional bias. The large overcall seen in the main text therefore depends on properties of the observed weight structure that the resampling generator retains and this generator does not; it is not a consequence of the nodal-strength sequence alone. The present analysis does not identify which specific weight-structure property is responsible. This is why the main text describes the 43% figure as specific to its reference ensemble rather than as a universal property of the reshuffle, re-selected specification. Note that the two strength-conditioned specifications sit at the nominal 5% here, far below the reshuffle, re-selected rate under the main-text generator, and neither shows directional bias.

## S6 Pseudo-network generation procedure

Inputs: the participant network $W$, its montage mask, and the number of rescaling iterations $T$ = 40. Output: a pseudo-network on the same mask whose nodal strengths match those of $W$ to numerical precision but whose edge weights are drawn afresh.

1. Draw an independent gamma-distributed value (shape 2, scale 1) for every permissible edge and symmetrise; call the result $W_0$.
2. Let $s^*$ be the nodal-strength vector of the observed network $W$.
3. For $t$ = 1 to $T$: compute the current strength vector $s$ of $W_{t-1}$; set $f_n = \frac{s^*_n}{s_n}$ for every node (with $f_n = 1$ where $s_n = 0$); multiply every permissible edge $(i, j)$ by $\sqrt{f_i f_j}$ to obtain $W_t$.
4. Return $W_T$.

Each iteration scales an edge by the geometric mean of its two end-nodes' correction factors, so the strength vector converges to $s^*$ while the weights retain the random start; the reshuffled and strength-conditioned comparison networks are then generated from the pseudo-network exactly as in the main analysis. Independent draws for the same participant have a mean pairwise edge-weight correlation of 0.25 and a strength-matching error of 0.00%. This algorithm is implemented in rand_sinkhorn in calib_randsink_release.py of the released code.

## S7 Permissible-edge count and surrogate-level hub re-selection

Because the permissible-edge set is uneven (Table S1), a node's expected strength under the global weight reshuffle is proportional to its number of permissible edges, so re-selecting hubs inside each reshuffled surrogate could favour the derivations with the most permissible edges regardless of the observed connectivity. To quantify this, 1,000 surrogates per participant were drawn under each null (eyes-open networks, $k$ = 5, fixed seeds), the five strongest nodes of every surrogate were recorded, and two quantities were computed: the frequency with which each derivation was re-selected as a hub, and the proportion of the participant's five observed hubs recovered by the surrogate hub set. Table S7 reports these quantities by derivation.

| **Bipolar derivation** | **Permissible edges** | **Observed hub frequency (%)** | **Re-selected, reshuffle (%)** | **Re-selected, strength-conditioned (%)** |
|---|---|---|---|---|
| **F8-F4** | 20 | 30.0 | 36.8 | 30.9 |
| **F7-F3** | 20 | 22.5 | 37.2 | 23.9 |
| **F4-C4** | 17 | 15.0 | 14.7 | 15.5 |
| **F3-C3** | 17 | 17.5 | 14.4 | 18.7 |
| **F4-FZ** | 18 | 0.0 | 21.0 | 0.0 |
| **F3-FZ** | 18 | 12.5 | 20.6 | 10.8 |
| **FZ-CZ** | 17 | 7.5 | 14.3 | 6.7 |
| **T4-C4** | 18 | 20.0 | 20.9 | 20.1 |
| **T3-C3** | 18 | 0.0 | 20.6 | 1.4 |
| **C4-CZ** | 16 | 27.5 | 10.1 | 25.3 |
| **C3-CZ** | 16 | 15.0 | 9.7 | 15.6 |
| **CZ-PZ** | 17 | 27.5 | 14.5 | 26.3 |
| **C4-P4** | 17 | 10.0 | 14.5 | 10.8 |
| **C3-P3** | 17 | 15.0 | 14.9 | 14.8 |
| **T4-T6** | 20 | 80.0 | 36.9 | 78.8 |
| **T3-T5** | 20 | 17.5 | 37.2 | 20.6 |
| **P4-PZ** | 18 | 15.0 | 20.8 | 13.9 |
| **P3-PZ** | 18 | 27.5 | 21.0 | 27.7 |
| **T6-O2** | 19 | 15.0 | 28.8 | 14.9 |
| **T5-O1** | 19 | 52.5 | 28.6 | 51.9 |
| **P4-O2** | 18 | 25.0 | 20.8 | 23.8 |
| **P3-O1** | 18 | 10.0 | 20.6 | 11.9 |
| **O1-O2** | 18 | 37.5 | 21.2 | 35.7 |

**Table S7:** Permissible-edge count, observed hub frequency, and surrogate-level hub re-selection by bipolar derivation (eyes-open networks, $k$ = 5, 1,000 surrogates per participant under each null). Observed hub frequency, the percentage of participants in whom the derivation is among the five strongest nodes; re-selection frequency, the percentage of surrogates in which it is among the five strongest nodes of the surrogate, averaged across participants. Under the global reshuffle, re-selection frequency is almost fully determined by the permissible-edge count (Spearman $\rho$ = 0.96): the four derivations with 20 permissible edges are re-selected in 37% of surrogates each, the two with 16 in 10%, and the surrogate hub set recovers on average 1.2 of the 5 observed hubs, similarly in the two groups (AD 26%, HC 24%). Under the strength-conditioned null, surrogates recover 4.8 of the 5 observed hubs and re-selection follows the observed hub frequency rather than the edge count ($\rho$ = 0.34), while the observed hubs themselves are only weakly related to edge count ($\rho$ = 0.33; T4-T6, the most frequent observed hub, is among the five strongest nodes in 80% of participants). The uneven montage therefore interacts with the global reshuffle through surrogate-level re-selection: the comparison coefficient of the mismatched specification is computed for a mask-determined node set rather than the observed one. Because this occurs similarly in both groups, it changes what the specification estimates rather than producing the group difference on its own.

## S8 Internal hub-strength fraction

A difference in the raw weighted rich-club coefficient could arise in two ways: the hub nodes could direct a greater share of their connectivity to one another (genuine preferential coupling), or the hub nodes could simply be stronger overall, carrying more weight on every edge including those between hubs. To separate these, the internal hub-strength fraction was computed for each participant as the proportion of the five hub nodes' total strength that falls on hub-to-hub edges. Table S8 reports this fraction alongside the total hub strength and the raw hub-to-hub edge weight (the numerator of the unnormalised coefficient).

| Measure | AD (mean ± SD) | HC (mean ± SD) | $p$ | $r$ |
|---|---|---|---|---|
| **Raw hub-to-hub edge weight** | 2.54 ± 0.79 | 2.03 ± 0.65 | 0.060 | -0.35 |
| **Total hub strength** | 19.5 ± 6.1 | 15.7 ± 2.9 | 0.050 | -0.36 |
| **Internal hub-strength fraction** | 0.266 ± 0.056 | 0.259 ± 0.063 | 0.756 | -0.06 |

**Table S8:** Decomposition of hub connectivity in the eyes-open networks ($k$ = 5). Values are group means ± standard deviation across participants; $p$, Mann-Whitney U test; $r$, rank-biserial effect size (a negative $r$ indicates higher values in the Alzheimer's disease group). The raw hub-to-hub edge weight is numerically higher in the AD group ($p$ = 0.060), which is the raw difference the reshuffle, re-selected specification reports as a rich-club effect. However, the total strength of the hub nodes is also higher in the AD group ($p$ = 0.050), while the internal hub-strength fraction, the share of that strength falling on hub-to-hub edges, does not differ between groups (0.27 versus 0.26, $p$ = 0.756). In other words, the analyses did not identify greater preferential coupling among the AD group's hubs; the hubs simply carry more weight overall, so their mutual edges are heavier in proportion. The raw hub-to-hub difference therefore reflects elevated hub strength rather than preferential coupling within the core, consistent with the absence of a group difference under the strength-conditioned, fixed-hub specification in the main analysis.

**S9 Relative core enrichment by edge class**

A second, independent check of whether the core is preferentially weighted in the AD group uses edge classes rather than the rich-club coefficient. Each edge was labelled by the roles of the nodes it connects: core-core (both nodes are hubs), core-periphery (one hub), or periphery-periphery (neither). To remove the overall strength difference between groups, each participant's edge weights were first normalised by their own network mean. Core enrichment was then quantified as the contrast between the core-core class and each other class (core-core minus core-periphery, and core-core minus periphery-periphery), and these contrasts were compared between groups. Significance was assessed by a label-permutation test (10,000 permutations of group labels).

| Core-enrichment contrast | AD | HC | $r$ | Permutation $p$ |
|---|---|---|---|---|
| **Core-core minus core-periphery** | 0.75 | 0.68 | -0.09 | 0.625 |
| **Core-core minus periphery-periphery** | 1.16 | 1.09 | -0.08 | 0.635 |

**Table S9:** Relative core enrichment by edge class in the eyes-open networks ($k$ = 5), after normalising each participant's edge weights by their network mean. Core enrichment is the contrast between the core-core edge class and each other class; values are the group mean contrasts. $r$, rank-biserial effect size (a negative $r$ indicates higher values in the Alzheimer's disease group); permutation $p$, from 10,000 permutations of the group labels. The core-core class is more heavily weighted than the other classes in both groups, as expected when hubs are strongly connected, but the degree of this enrichment does not differ between groups (both permutation $p \geq 0.5$). Because the weights were normalised within participants before the contrast, this comparison is insensitive to the overall difference in connectivity between groups, and isolates whether weight is concentrated into the core specifically. No such group difference is present, which agrees with the internal hub-strength fraction (Table S8) and with the absence of a group difference under the strength-conditioned, fixed-hub specification in the main analysis: the AD group's higher raw hub-to-hub weight reflects stronger hubs overall rather than preferential coupling within the core.

## S10 Sensitivity to the hub definition

The main analysis defines hubs as the five nodes of highest strength. Because the permissible-edge set is uneven (Table S1), a node's strength depends partly on how many permissible edges it has, which could influence which nodes are selected as hubs. To check that the results do not hinge on this, hubs were redefined in two further ways, by mean permissible-edge weight (strength divided by the number of permissible edges, removing the edge-count component) and by strength residualised on the number of permissible edges (regressing edge count out of strength), and both the preferred and the mismatched pipeline were re-run under each definition with 300 nulls per participant and independent seeds. The strength-defined rows therefore reproduce the Table 1 specifications at lower precision and differ from them only by Monte Carlo error. Table S10 reports the eyes-open group comparison at $k = 5$ under each definition.

***Preferred specification (strength-conditioned null, fixed observed hubs)***

| Hub definition | AD | HC | $p$ | $r$ |
|---|---|---|---|---|
| **Strength** | 1.129 | 1.094 | 0.490 | -0.13 |
| **Mean permissible-edge weight** | 1.180 | 1.134 | 0.507 | -0.12 |
| **Strength residualised on edge count** | 1.184 | 1.171 | 0.968 | +0.01 |

***Mismatched specification (weight-reshuffle null, re-selected hubs)***

| Hub definition | AD | HC | $p$ | $r$ |
|---|---|---|---|---|
| **Strength** | **1.093** | **0.928** | **0.029** | **-0.41** |
| **Mean permissible-edge weight** | **1.104** | **0.938** | **0.022** | **-0.42** |
| **Strength residualised on edge count** | 1.127 | 0.976 | 0.126 | -0.28 |

**Table S10:** Eyes-open group comparison of the normalised weighted rich-club coefficient ($k = 5$) under three hub definitions, for the preferred specification and the mismatched specification. AD and HC, group mean normalised coefficient; $p$, Mann-Whitney U test; $r$, rank-biserial effect size (a negative $r$ indicates higher values in the Alzheimer's disease group); bold rows are significant at $p < 0.05$. The preferred specification shows no group difference under any of the three hub definitions (all $p > 0.05$), so the null result does not hinge on defining hubs by strength. The mismatched specification produces a group difference under the strength and mean-edge-weight definitions, as in the main analysis, but this difference weakens and is no longer significant when hubs are defined by strength residualised on the number of permissible edges ($p$ rises from 0.03 to 0.13). Removing the edge-count component from the hub definition thus attenuates the analysis-dependent apparent effect, indicating that part of what the mismatched specification detects is tied to the uneven permissible-edge structure of the montage rather than to preferential coupling. This reinforces the main conclusion: the apparent effect is a product of the analysis specification, here interacting with the montage, rather than a robust group difference.

**S11 Global connectivity by group and recording state**

The apparent rich-club effect was found in the eyes-open state. Because overall connectivity also differs between the eyes-open and eyes-closed states, it is worth checking whether the group difference is specific to one state or whether both groups simply respond to eye closure in the same way. Mean global connectivity (the average of all permissible edge weights) was compared between groups in each state, and the within-participant change between states (eyes-open minus eyes-closed) was compared between groups to test for a group-by-state interaction. Table S11 reports these comparisons.

| Comparison | AD (mean ± SD) | HC (mean ± SD) | $p$ | $r$ |
|---|---|---|---|---|
| **Eyes-open connectivity** | **0.160 ± 0.054** | **0.130 ± 0.023** | **0.041** | **-0.38** |
| **Eyes-closed connectivity** | 0.211 ± 0.092 | 0.181 ± 0.061 | 0.473 | -0.14 |
| **State change (eyes-open − eyes-closed)** | -0.050 ± 0.054 | -0.051 ± 0.063 | 0.776 | +0.06 |

**Table S11:** Mean global connectivity by group and recording state. Values are group means ± standard deviation across participants; $p$, Mann-Whitney U test; $r$, rank-biserial effect size (a negative $r$ indicates higher values in the Alzheimer's disease group); bold rows are significant at $p < 0.05$. Global connectivity is higher in the AD group in the eyes-open state ($p = 0.041$) but not in the eyes-closed state ($p = 0.473$). Critically, the within-participant change in connectivity between states is almost identical in the two groups (AD -0.050, HC -0.051; group-by-state interaction $p = 0.776$): both groups reduce connectivity by the same amount on eye opening. The group difference is therefore a difference in overall connectivity level that is detectable in the eyes-open state, not a difference in how connectivity responds to eye state. For this reason the eyes-open effect is not interpreted as state-dependent in the main text. This higher eyes-open connectivity may contribute to greater raw hub strength but does not, by itself, establish preferential hub coupling or explain the re-selected rich-club contrast.